\documentclass{Interspeech2024}
\usepackage{multirow,adjustbox,graphicx,subcaption,hyperref}
\usepackage{cite}
\usepackage{hyperref}
\hypersetup{
    colorlinks=true,
    linkcolor=red,
    urlcolor=magenta,
    }

\interspeechcameraready

\title{SAML: Speaker Adaptive Mixture of LoRA Experts for End-to-End ASR}

\name[affiliation={1}]{Qiuming}{Zhao}
\name[affiliation={2}]{Guangzhi}{Sun}
\name[affiliation={1}]{Chao}{Zhang}
\name[affiliation={1}]{Mingxing}{Xu}
\name[affiliation={1*}]{Thomas Fang}{Zheng}

\address{
  $^1$Tsinghua University, China \\
  $^2$University of Cambridge, United Kingdom\thanks{*Correspondence}}
\email{zqm23@mails.tsinghua.edu.cn, gs534@cam.ac.uk, \{cz277,xumx,fzheng\}@tsinghua.edu.cn}


\begin{document}

\maketitle

\begin{abstract}
Mixture-of-experts (MoE) models have achieved excellent results in many tasks. However, conventional MoE models are often very large, making them challenging to deploy on resource-constrained edge devices. In this paper, we propose a novel speaker adaptive mixture of LoRA experts (SAML) approach, which uses low-rank adaptation (LoRA) modules as experts to reduce the number of trainable parameters in MoE. Specifically, SAML is applied to the quantised and personalised end-to-end automatic speech recognition models, which combines test-time speaker adaptation to improve the performance of heavily compressed models in speaker-specific scenarios. Experiments have been performed on the LibriSpeech and the TED-LIUM 3 corpora. Remarkably, with a 7x reduction in model size, 29.1\% and 31.1\% relative word error rate reductions were achieved on the quantised Whisper model and Conformer-based attention-based encoder-decoder ASR model respectively, comparing to the original full precision models.
\end{abstract}

\section{Introduction}
Transformer or Conformer-based end-to-end neural network models have achieved state-of-the-art performance in Automatic Speech Recognition (ASR) tasks \cite{radford2023robust,gulati2020conformer,baevski2020wav2vec}. While these end-to-end ASR models are becoming more capable and generalisable via large-scale training, the model sizes increase significantly, motivating efficient training approaches to be explored when there is limited task-specific data \cite{whisperbiasing,zhao2023enhancing}. Speaker adaptation is a typical task of this kind, where a generic ASR system is to be adapted to perform better for a certain speaker by providing only a handful of annotated speech data from that speaker. Previous work has explored using the low-rank adaptation (LoRA) \cite{hu2021lora} in combination with model quantisation for speaker adaptation \cite{zhao2023enhancing}. However, a single static set of adaptation parameters to handle speaker variability may yield a sub-optimal solution in speaker adaptation \cite{tan2015cluster,yu2006discriminative,kuhn2000rapid}, necessitating the use of a dynamic network design for enhanced adaptation performance, such as Mixture-of-Experts approaches.


Mixture-of-Experts (MoE) Transformer-based models have received extensive research attention in fields such as natural language processing \cite{fedus2022switch,shazeer2017outrageously,du2022glam}, speech processing \cite{you2021speechmoe,perez2020aphasic}, and computer vision \cite{riquelme2021scaling,chen2023adamv}. Concretely, the MoE is a family of neural network architectures that enables conditional computation through multiple experts that are activated based on a gating network, referred to as the router. This mechanism effectively enhances model representation power and expands model capacity. Furthermore, sparse MoE \cite{fedus2022switch} activates only one sub-network for each input data, improving training and inference efficiency. However, these advantages come at the cost of dramatic increases in model size, which not only increases operational costs on the server but also presents significant challenges in deploying them on resource-constrained edge devices.


To combine the advantages of both LoRA and MoE for speaker adaptation, this paper proposes the Speaker Adaptive Mixture of LoRA experts (SAML) approach that adopts LoRA modules as experts. As a LoRA-based approach, SAML significantly releases the burden of the number of parameters in MoE. Specifically, SAML is integrated into the personalisation for a quantised model (PQM) framework \cite{zhao2023enhancing}. In PQM, block-wise NormalFloat4 (NF4) quantisation \cite{dettmers2023qlora} is adopted to achieve model compression, which incurs a smaller performance loss compared to conventional uniform quantisation. 
The SAML-based speaker adaptation is applied on top of the quantised models to compensate for the degradation due to quantisation. 
This is based on the fact that the edge devices to deploy quantised models are often personalised. For these devices, such as personalised voice assistants or smart door locks, improving performance for the target speaker is the critical objective rather than the performance concerning other speakers.

The SAML approach was implemented for the Conformer attention-based encoder-decoder (AED) model and the Whisper model as two examples of end-to-end ASR models in this paper. Experiments performed on the LibriSpeech and TED-LIUM 3 datasets demonstrated that, with nearly a 7$\times$ compression of the model in the PQM framework, using SAML achieves a relative WER reduction of 29.1\% and 31.1\% on quantised Whisper and Conformer AED models respectively, compared to the original full-precision models. The main contribution of this paper can be summarised as follows.
\begin{itemize}
\setlength\itemsep{0.1em}
    \item We propose SAML as the first LoRA-based MoE approach for speaker adaptation.
    \item We integrate SAML into the PQM framework to further compensate for the degradation incurred in the quantisation.
    \item SAML has been validated on both Conformer AED and Whisper models across two datasets, with superior performance over single LoRA-based adaptation methods.
\end{itemize}


\begin{figure*}[t]
  \centering
  \includegraphics[width=0.9\linewidth]{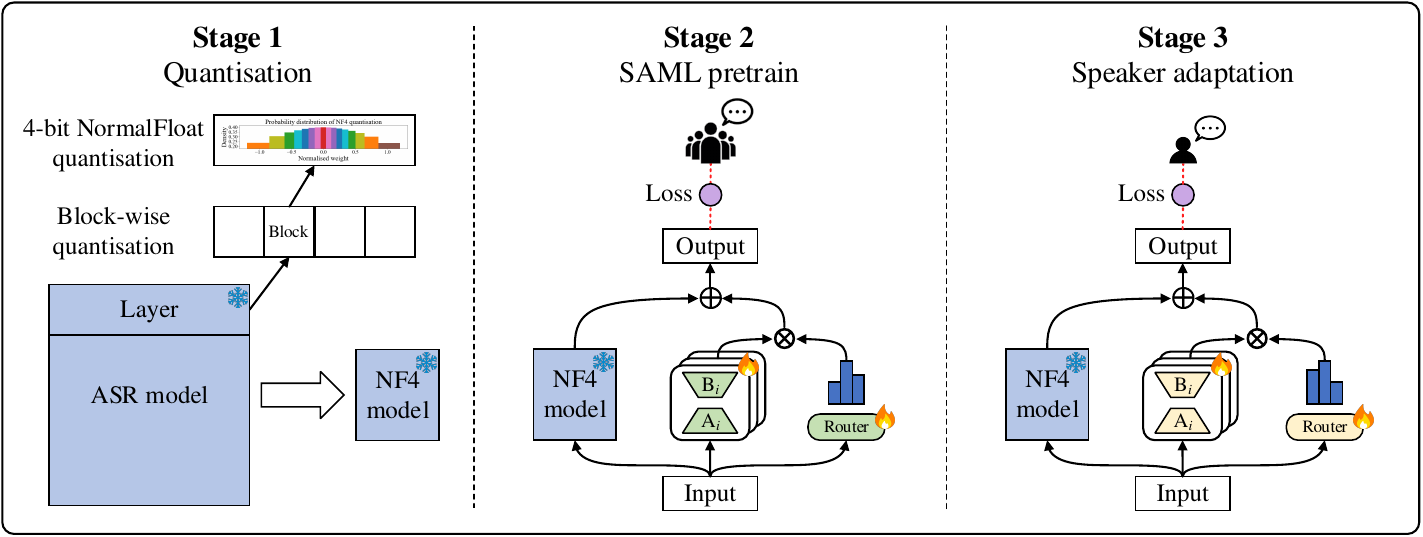}
  \vspace{-0.1cm}
  \caption{Overview of the SAML integrated into PQM framework.}
  \label{fig:Overview}
  \vspace{-0.3cm}
\end{figure*}

\section{Related work}
\label{sec:related work}

\subsection{Mixture-of-Experts}
MoE is an effective method to expand model capacity. Recently, some studies investigated the scale properties \cite{zoph2022st,rajbhandari2022deepspeed,clark2022unified} of MoE models.
Routing algorithms have also been studied extensively. Classic routing algorithms include soft routing \cite{jacobs1991adaptive,puigcerver2023sparse} and top-k routing \cite{fedus2022switch,shazeer2017outrageously}, among others \cite{roller2021hash,lewis2021base}.
Moreover, MoE models have been applied to multimodal \cite{mustafa2022multimodal,akbari2024alternating} and multitask \cite{ma2018modeling,chen2023mod} learning, illustrating their adaptability across diverse domains. To improve the deployment of MoE models, several studies have applied techniques such as quantisation \cite{frantar2023qmoe}, pruning \cite{kim2021scalable}, and distillation \cite{fedus2022switch} to reduce the size and memory footprint of MoE models. Our objectives align with theirs, but we primarily propose the SAML approach to achieve a lightweight MoE model.

\subsection{Speaker adaptation}
\label{ssec:speaker adaptation}
The objective of speaker adaptation is to minimize the mismatch between speakers in training and testing conditions.
Embedding-based adaptation methods map speakers into a continuous space using techniques like i-vectors \cite{sari2020unsupervised} or neural network bottlenecks \cite{yue2020autoencoder}.
Model-based adaptation methods \cite{swietojanski2016learning,zhang2016dnn,yu2013kl} adjust the model structure and parameters to individual speakers. 
Recent LoRA-based adaptation methods \cite{zhao2023enhancing,hsieh2022adapter,chen2024hyporadise} have been widely applied in many tasks. Compared to full fine-tuning, LoRA adjusts only the low-rank subspace parameters of the model, thereby achieving higher computational efficiency and lower costs for computation and storage.



\section{Methodology}
\label{sec:methodology}
\subsection{Preliminaries}

\subsubsection{Mixture-of-Experts}
An MoE layer consists of a router network $G$ and a set of $n$ expert networks $E_{1}$,\ldots,$E_{n}$. The output $h$ of the MoE layer can be expressed as follows:
\begin{equation}
h = E_\text{mix} = \sum_{i=1}^{n} G(x)_i E_i(x)
\end{equation}
For soft routing, $G(\cdot)$ calculates scores for each expert based on the input $x$:
\begin{equation}
G(x) = \text{Softmax}(W_g \cdot x)
\end{equation}
where $W_g$ is the weight matrix of the router $G$.

\subsubsection{Low-Rank Adaptation}
LoRA \cite{hu2021lora} is a parameter-efficient fine-tuning method. For the pretrained model with weight matrix \( W_0 \in \mathbb{R}^{d \times k} \), its forward pass yields:
\begin{equation}
h = W_0 x + \Delta W x = W_0 x + \frac{\alpha}{r} B A x
\end{equation}
where $\alpha$ is a scaling factor that adjusts the magnitude of the changes to the original $W_0$ made by the LoRA module, \( B \in \mathbb{R}^{d \times r} \), \( A \in \mathbb{R}^{r \times k} \), and the rank \( r \ll \min(d, k) \).

\subsection{Speaker Adaptive Mixture of LoRA Experts}
\begin{figure}[t]
  \centering
  \includegraphics[width=0.8\linewidth]{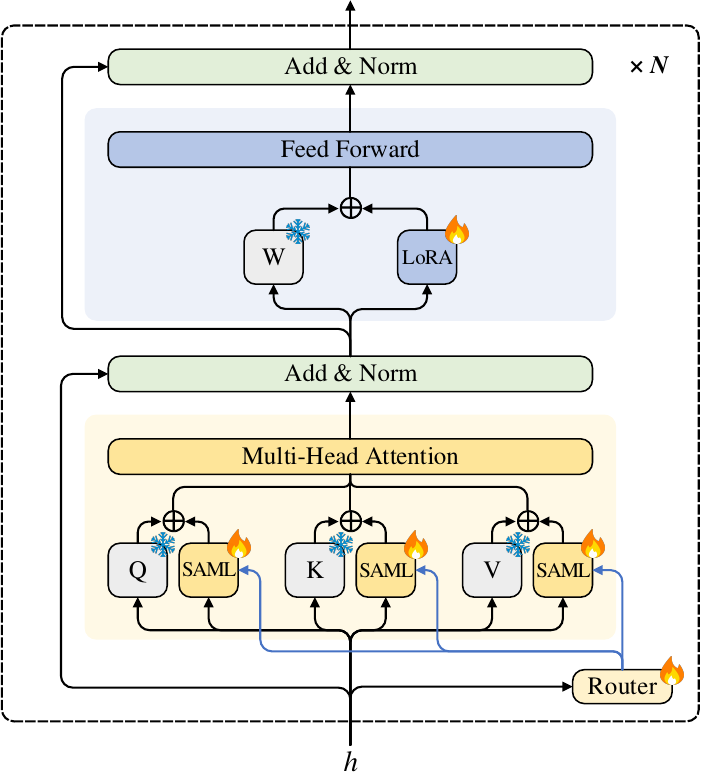}
  \caption{The SAML architecture. Each attention layer is replaced with the SAML layer, and a LoRA module is added to each feed-forward layer.
  }
  \label{fig:SAML}
  \vspace{-0.3cm}
\end{figure}

MoE models dynamically select and weigh experts based on input data through a dynamic routing mechanism, significantly enhancing the model representation power and scaling up the model capacity with only a minor computation overhead. 
However, previous works \cite{fedus2022switch,shazeer2017outrageously} adopted dense feed-forward networks as experts, leading to dramatic increases in model size.
Consequently, we propose the SAML, which adopts parameter-efficient LoRA modules as experts, significantly reducing the parameter burden in MoE models.
The specific details of the SAML architecture are shown in Figure \ref{fig:SAML}.

The output $h$ of the SAML layer is:
\begin{equation}
h = W_0 x + \Delta W x = W_0 x + E_\text{mix}
\end{equation}
where $E_\text{mix}$ is the mixture of LoRA experts by soft routing:
\begin{equation}
E_\text{mix} = \frac{\alpha}{r} (\sum_{i=1}^{n} G(x)_i B_i) (\sum_{i=1}^{n} G(x)_i A_i) x
\end{equation}
Compared to the multiplication of $A_i$ and $B_i$ before addition, we adopted adding $A_i$ and $B_i$ before their multiplication which is more efficient in terms of GPU memory because it circumvents multiple matrix multiplication across different LoRA modules.
Additionally, we quantise both the router and the experts to further reduce the model size.

\subsection{SAML integrated into PQM framework}
The SAML integrated into the PQM framework is illustrated in Figure \ref{fig:Overview}, which is divided into three stages. In stage 1, we apply block-wise NF4 quantisation to the base model's primary weight parameters. In stage 2, we pretrain the router and the LoRA experts' parameters using data from a large number of speakers, providing a more robust starting point for subsequent speaker adaptation. In stage 3, we perform SAML-based speaker adaptation on speaker-specific data.


The block-wise NF4 quantisation is adopted in the PQM framework. While standard floating point quantisation applies the same set of quantisation bins to all weight matrices, the dynamic range of parameter values is not taken into account, resulting in heavily unbalanced quantisation bins. NF4, on the contrary, ensures each bin has an equal number of values by estimating the quantile of the input matrices using the empirical cumulative normal distribution. This leveraged the fact that the parameters of a weight matrix, in general, follow a normal distribution \cite{dettmers2023qlora}.

To reduce the influence of extreme values in weight matrices (i.e. outliers) on the maximum absolute value normalisation, block-wise quantisation is applied which divides the weight matrices into small blocks and quantises each block with separate normalisation factors.
In this way, outliers in the input tensor are confined to individual blocks, reducing their overall impact on quantisation. As a result, block-wise quantisation allows for individual normalisation factors for each block, resulting in a more fine-grained overall quantisation.

Although the target speaker data is always limited, in reality, the target domain data of other speakers is usually available. Therefore, PQM leverages those data to find a better initialisation point for SAML weights before performing speaker adaptation, referred to as SAML pretraining.
In speaker adaptation, the base model is frozen, and only the router and the LoRA experts' parameters corresponding to each speaker are updated.

\section{Experimental setup}
\label{sec:experiments}
\subsection{Data}


{\bf LibriSpeech} is an English audiobook dataset. We selected 5 male speakers and 5 female speakers with the largest number of utterances from train-clean-360 as speaker adaptation data. Each speaker contributes approximately 150 utterances, resulting in a total speech duration of roughly 25 minutes. For SAML pretraining, the train-clean-100 set was used which does not have any speaker overlap with the selected speakers.

{\bf TED-LIUM 3} (TL3) is a TED talks dataset. We selected 16 speakers from the test set as speaker adaptation data. On average, each speaker has 161 utterances (14 minutes).

Speaker adaptation data for LibriSpeech and TL3 was divided randomly, where 2/5 was divided into the train set, 1/5 was divided into the dev set, and 2/5 was divided into the test set. On average, each speaker has 6-10 minutes of training data, while the dev and test data remains constant across all experiments. We denote the partitioned test sets as \textit{LibriSpeech-SA} and \textit{TL3-SA} respectively in the results. 
\footnote{Code and data partition: \href{https://github.com/qmgzhao/SAML.git}{https://github.com/qmgzhao/SAML.git}}

\subsection{Model and training specifications}
\label{ssec:baseline}

To verify the effectiveness of SAML, we use the Whisper and Conformer AED models as two widely used models as examples.

{\bf Whisper} is a Transformer-based AED model released by OpenAI trained on 680k hours of audio. The base.en model with a full model size of 278MB was used. The encoder has 6 Transformer blocks with 2048 hidden dimensions, and the output size is 512. The decoder has 6 Transformer blocks with 2048 hidden dimensions. The Transformer-related weight matrices are all 512 by 512 dimensional. Feature processing and model training followed \cite{radford2023robust,whisperbiasing,zhao2023enhancing}.

{\bf Conformer AED} is a hybrid CTC/attention-based encoder-decoder model, whose FP32 model size is about 131MB. The training follows ESPnet \cite{espnet} with 0.3 CTC weight and 80-dim FBank features. The Conformer encoder has 12 blocks with 1024 hidden dimensions. The decoder uses a 6-block Transformer architecture with 2048-dim linear units. The Transformer-related weight matrices are all 256 by 256.

During the pretraining and adaptation stages, we conduct joint interleaved training of the router and experts. 
The default number of experts for all layers is uniformly set to 10, with the LoRA rank set to 1 for Whisper and 4 for Conformer. Furthermore, each group of experts has initialised with the LoRA parameters pretrained on a single speaker data from the train-clean-100 set. Models are evaluated using WER averaged across all utterances from the test set speakers.

\section{Evaluation results and analysis}
\label{sec:results}
First, we apply block-wise NF4 quantisation to primary weight parameters of the model, including linear, convolution, and embedding layers, resulting in a 7$\times$ reduction in the size of both Whisper and Conformer models. For detailed WER and model compression ratios of systems after quantising different parts, please refer to \cite{zhao2023enhancing}. Furthermore, WER increased by 1.20\% for Whisper and only 0.34\% for Conformer upon NF4 quantisation. This suggests that models trained on smaller datasets are more robust to the quantisation noises under NF4 quantisation.

\begin{table}[t]
  \caption{WER on the LibriSpeech-SA and TL3-SA using quantised Whisper models. Parameter size lists the size (MB) of total parameters and trainable parameters (in parentheses). FFT refers to full fine-tuning which trains all model parameters. LoRA refers to a single LoRA and SAML refers to the mixture of LoRA experts.}
  \label{tab:whisper}
  \centering
  \setlength{\tabcolsep}{1.5pt}
  \begin{adjustbox}{width=\columnwidth}
  \begin{tabular}{lccc}
    \toprule
    \multicolumn{1}{l}{\multirow{2}{*}{\textbf{System}}} & \multicolumn{1}{c}{\multirow{2}{*}{\textbf{Param. Size}}} & \multicolumn{2}{c}{\textbf{WER(\%)}} \\
     & & LibriSpeech-SA & TL3-SA \\
    \midrule
    Whisper-FP32                   & 277.8         & 10.02 & 5.93 \\
    Whisper-NF4                    & 38.3          & 11.22 & 7.71 \\
    Whisper-FFT-FP32               & 277.8 (277.8) & 9.05  & 6.43 \\
    Whisper-FFT-NF4                & 38.3 (38.3)   & 10.59 & 7.20 \\
    Whisper-LoRA-FP32              & 38.6 (0.3)    & 8.48  & 6.87 \\
    Whisper-LoRA-NF4               & 38.4 (0.1)    & 8.51  & 6.72 \\
    \midrule
    Whisper-SAML-pretrain-FP32   & 44.6 (6.3)    & 7.90  & 5.49 \\
    Whisper-SAML-pretrain-NF4    & 39.5 (1.2)    & 7.99  & 5.30 \\
    Whisper-SAML-adaptation-FP32 & 44.6 (6.3)    & 6.94  & 4.95 \\
    Whisper-SAML-adaptation-NF4  & 39.5 (1.2)    & \textbf{7.10}  & \textbf{4.72} \\
    \bottomrule
  \end{tabular}
  \end{adjustbox}
\end{table}
Table \ref{tab:whisper} shows the performance of the SAML approach on the Whisper base.en model. Compared to Whisper-NF4, the WER reduction achieved by fine-tuning all model parameters at full precision on pretraining data and target speaker data was largely reduced after model quantisation. As a result, the Whisper-FFT-NF4 model only achieved around 5.6\% relative WER reduction on LibriSpeech-SA and 6.6\% relative WER reduction on TL3-SA tasks. 
Due to LoRA updating only a small amount of the low-rank subspace parameters, which enhances its robustness to quantisation, there is almost no performance degradation after quantisation. LoRA at NF4 precision achieved 24.2\% and 12.8\% relative WER reductions respectively. 
When SAML was applied to the model, only with pretraining, the performance already surpassed that of both FFT and LoRA. Moreover, with speaker adaptation, the improvements at NF4 precision were further enlarged, resulting in 36.7\% and 38.8\% relative WER reductions on LibriSpeech-SA and TL3-SA sets respectively. Note that the SAML pretraining for TL3-SA was cross-data, as the pretraining was done on the LibriSpeech clean-100 training set while directly applied to the TL3-SA data for speaker adaptation. This underscores the effectiveness of the pretrained SAML approach.

\begin{table}[t]
  \caption{WER on the LibriSpeech-SA using quantised Conformer models. FFT, LoRA and SAML follow the same definition as Table \ref{tab:whisper}.}
  \label{tab:conformer}
  \centering
  \begin{adjustbox}{width=0.95\columnwidth}
  \begin{tabular}{@{\hspace{2pt}}lcc}
    \toprule
    \textbf{System} & \textbf{Param. Size} & \textbf{WER(\%)} \\
    \midrule
    Conformer-FP32                   & 130.9         & 12.43 \\
    Conformer-NF4                    & 19.1          & 12.77 \\
    Conformer-FFT-FP32               & 130.9 (130.9) & 8.46 \\
    Conformer-FFT-NF4                & 19.1 (19.1)   & 10.52 \\
    Conformer-LoRA-FP32              & 19.9 (0.8)    & 9.53 \\
    Conformer-LoRA-NF4               & 19.3 (0.2)    & 9.54 \\
    \midrule
    Conformer-SAML-pretrain-FP32   & 28.0 (8.9)    & 9.99 \\
    Conformer-SAML-pretrain-NF4    & 20.8 (1.7)    & 9.98 \\
    Conformer-SAML-adaptation-FP32 & 28.0 (8.9)    & 8.55 \\
    Conformer-SAML-adaptation-NF4  & 20.8 (1.7)    & \textbf{8.56} \\
    \bottomrule
  \end{tabular}
  \end{adjustbox}
  \vspace{-0.3cm}
\end{table}
The same set of experiments was also performed for the Conformer model as shown in Table \ref{tab:conformer}. Note that as the Conformer AED is trained on train-clean-100 already, we selected 250 speakers from LibriSpeech train-clean-360 for SAML pretraining. As before, the Conformer-SAML-pretrain-NF4 achieved a 21.8\% relative WER reduction compared to the Conformer-NF4. The Conformer-SAML-adaptation-NF4 model achieved a further WER reduction, resulting in a relative 33.0\% WER reduction.

\begin{table}[t]
  \caption{WER on the LibriSpeech-SA using Whisper-SAML-pretrain-FP32 with different numbers of experts.}
  \label{tab:experts number}
  \centering
  \begin{adjustbox}{width=0.6\columnwidth}
  \begin{tabular}{lcc}
    \toprule
    \textbf{System} & \textbf{WER(\%)} \\
    \midrule
    Whisper-SAML-5experts  & 8.02 \\
    Whisper-SAML-10experts & 7.90 \\
    Whisper-SAML-15experts & 7.85 \\
    Whisper-SAML-20experts & 7.82 \\
    \bottomrule
  \end{tabular}
  \end{adjustbox}
\end{table}
Next, Table \ref{tab:experts number} shows the performance of the Whisper-SAML-pretrain-FP32 model with different numbers of experts. 
The results demonstrate that performance consistently improves with an increasing number of experts, though at a diminishing rate for further increases.

\begin{table}[t]
  \caption{WER on the LibriSpeech-SA using Whisper-SAML-pretrain-FP32 with MoE pruning.}
  \label{tab:MoE pruning}
  \centering
  \begin{adjustbox}{width=0.9\columnwidth}
  \begin{tabular}{lcc}
    \toprule
    \textbf{System} & \textbf{WER(\%)} \\
    \midrule
    complete                                    & 7.90 \\
    pruning (delete non-collapsed experts \& router) & 7.90 \\
    keep top1 expert \& router           & 8.27 \\
    keep top1 expert            & 15.83 \\
    \bottomrule
  \end{tabular}
  \end{adjustbox}
  \vspace{-0.3cm}
\end{table}
In experiments, we observed that some SAML layers exhibit issues of load imbalance and model collapse \cite{chen2022towards}, with severe reliance on or even collapse into a single expert. We suggest that the collapsed layers might not require the complexity of the MoE architecture, since a single expert seems capable of handling their tasks. Therefore, we prune the collapsed layers by deleting all non-collapsed experts and the router, resulting in each collapsed layer degenerating into a single LoRA layer. 
Table \ref{tab:MoE pruning} shows the results of MoE pruning. Line 2 indicates that performance is lossless after MoE pruning.
Moreover, for layers with load imbalance, we also attempt to only keep the top1 expert and the router. Line 3 demonstrates that keeping the top1 expert and the router results in only a slight decrease in performance.
Line 4 shows that further deleting the router leads to a sharp decline in performance. 
This indicates that merely dynamic scaling on a single LoRA can yield significant effects.

\begin{figure}[h]
  \centering
  \vspace{-0.3cm}
  \begin{subfigure}[b]{0.8\linewidth}
    \centering
    \includegraphics[width=0.45\linewidth]{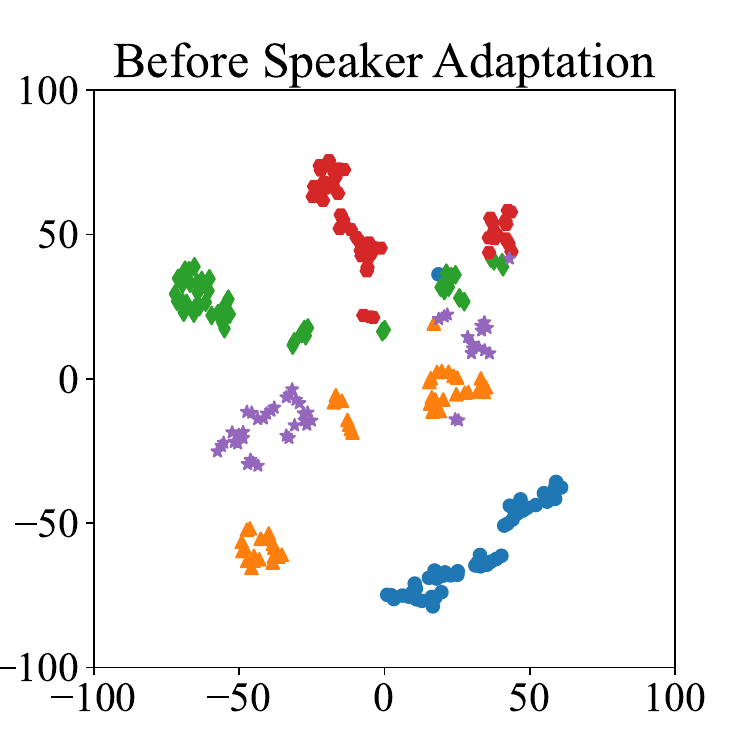}
    \includegraphics[width=0.45\linewidth]{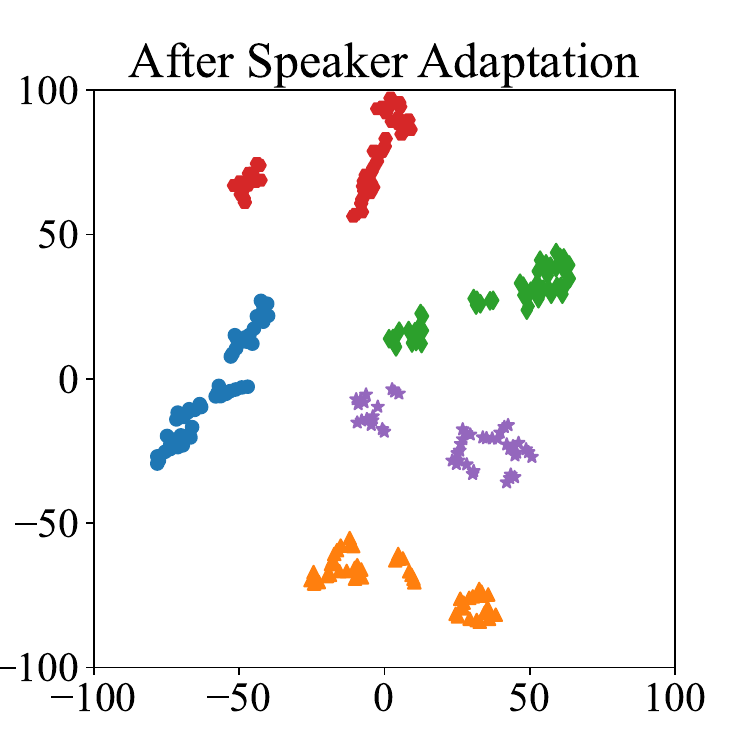}
    \vspace{-0.2cm}
    \caption{SAML}
  \end{subfigure}

  \begin{subfigure}[b]{0.8\linewidth}
    \centering
    \includegraphics[width=0.45\linewidth]{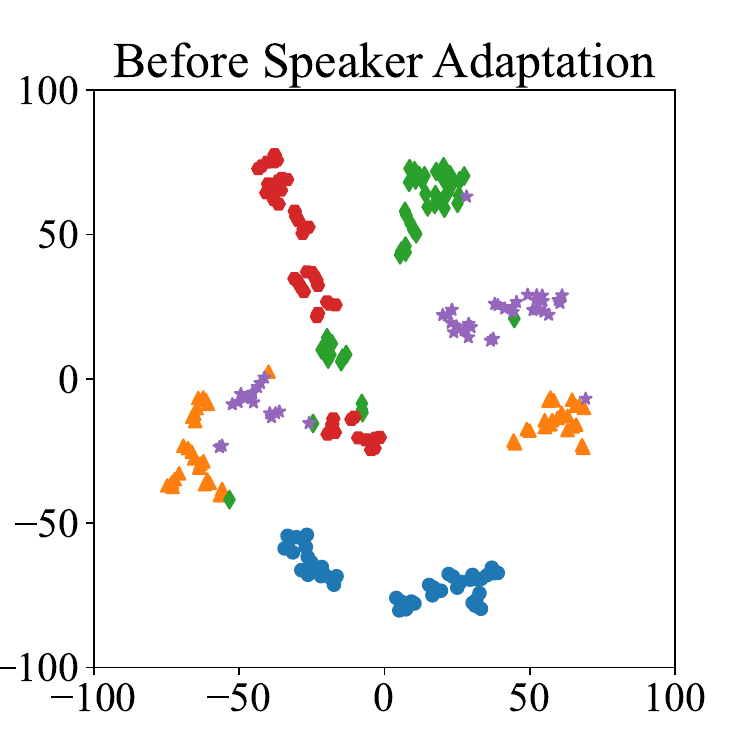}
    \includegraphics[width=0.45\linewidth]{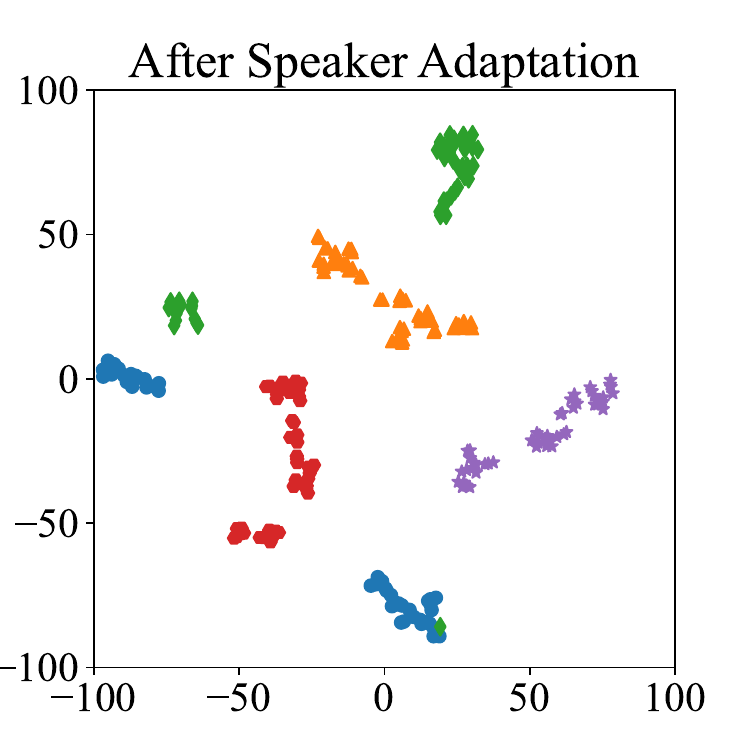}
    \vspace{-0.2cm}
    \caption{Single LoRA}
  \end{subfigure}

  \vspace{-0.2cm}
  \caption{t-SNE visualisation of Whisper-SAML and Whisper-LoRA encoder outputs, with different colours for each speaker.}
  \label{fig:speaker adaptation}
\end{figure}
Finally, the t-SNE visualisation results of speaker adaptation are displayed in Figure \ref{fig:speaker adaptation}. 
As shown, after speaker adaptation, both SAML and single LoRA have effectively captured speaker features.
Furthermore, for each speaker cluster, SAML achieves clearer separation, indicating that the experts in SAML provided better representation that enhanced speaker adaptive capabilities compared to the single LoRA.

\section{Conclusions}
\label{sec:conclusions}
This paper proposes the SAML approach and integrates it into the PQM framework. SAML, which uses LoRA modules as experts, is applied to both the Conformer-based AED and the Whisper ASR models. Experiments on LibriSpeech and TL3 datasets showed that SAML can largely reduce the WERs of the quantised models. Compared to the original full precision models, using SAML, 29.1\% and 31.1\% relative WER reductions were achieved on quantised Whisper and Conformer-based AED models respectively.

\newpage

\bibliographystyle{IEEEtran}
\bibliography{mybib}

\end{document}